\documentclass{aa}
\usepackage{astron,times,graphics}
\begin{document}
\thesaurus{07(08.16.4; 08.03.4; 08.12.1; 13.09.6)}

\title{The nature of the circumstellar CO$_2$ emission from M giants\thanks{Based 
on observations with ISO, an ESA project with instruments funded by ESA Member 
States (especially the PI countries: France, Germany, the Netherlands and the 
United Kingdom) and with the participation of ISAS and NASA. The SWS is a 
joint project of SRON and MPE.}}

\author{N.~Ryde\inst{1}  \and K.~Eriksson\inst{1} \and B. Gustafsson \inst{1} 
} 

\institute{Uppsala Astronomical Observatory, Box 515, S-751 20, Uppsala, Sweden
}
\offprints{N.~Ryde (ryde@astro.uu.se)}

\date{Received 21 July 1998; accepted 28 September 1998}

\maketitle

\begin{abstract}
The $13-16\mbox{$\mu$m}$ region observed by the Infrared Space Observatory (ISO) of several ABG stars are discussed.
We present and analyse spectra of seven M giants which show carbon dioxide features. To explain the features of the bands, we suggest they originate from two different circumstellar layers, one being a warm and high density layer close to the star, possibly making the $15\mbox{$\mu$m}$ band optically thick, and the other being  a large, cold and optically thin layer extending far out in the wind. 
This could explain the difference in temperatures of the different bands found in the analysis of the spectra and the number of molecules needed for the emission.

It is demonstrated that in spite of the bands probably not being formed in vibrational LTE, the temperatures can be estimated from the widths of the bands.

\end{abstract}

\section{Introduction}

Asymptotic Giant Branch (AGB) stars are considered to be important
contributors of heavy elements to the interstellar medium through their 
slow but massive stellar winds, and they thereby play an important role in the chemical evolution of galaxies, see e.g. Gustafsson \& Ryde \cite*{bg:96}. The 
physics of this mass-loss and the processes in winds  of AGB stars, such as the nucleation of dust grains and the acceleration of the wind, are, however,
not well understood. In trying to gain some insight into the 
physical and chemical conditions of the wind region and some understanding
of the mass-loss mechanism, the Infrared Space Observatory (ISO) has been used to observe the outer atmospheres and the circumstellar regions of a sample of AGB stars by observing spectra of 
diagnostic molecular bands. As one part in this 
project we have studied carbon dioxide in the oxygen-rich, semi-regular asymptotic giant branch 
stars 
\object{R Crt}, \object{R Dor}, \object{R Cas}, \object{RS CrA}, \object{TX Cam}, \object{V1111 Oph}, and \object{V656 Cas}. 

A dozen triatomic molecules have been detected in AGB stars to date, mainly at radio wave-lengths (cf. Olofsson et al., 1997\nocite{ho_mol}). From ground-based telescopes gas-phase CO$_2$ has been impossible to detect due to
the blocking by the terrestrial atmosphere. Only satellite observations provide an opportunity to observe astrophysical CO$_2$. 

The first report on CO$_2$ emission from M giants was given in Ryde et al. (1997), where it was suggested that the emission feature at $14.98 \mbox{$\mu$m}$ in \object{R Crateris} was due to the 
$00^00\, \Sigma_g^{ \,+}-01^10 \,\Pi_u$ 
Q-branch rotation-vibration transitions of circumstellar CO$_2$. Based on more data for six M giants, Justtanont et al. (1998)\nocite{jus} could prove unequivocally that this feature and others originate from circumstellar gaseous CO$_2$. On the basis of a preliminary modelling of the emission lines, Justtanont et al. \cite*{jus}  and Cami et al. \cite*{cami} suggest that the emission originates from in a region, in which molecules are formed, which extends a few stellar radii out from the star. From a thermal model they conclude that the gaseous CO$_2$ has a temperature of more than $650\mbox{K}$. 

Here, we present new ISO observations of the $13-15 \mbox{$\mu$m}$ region of several M giants.
CO$_2$ emission is detected to various degrees in these stars.

\section{Observations and reductions}

\begin{figure*}
  \resizebox{\hsize}{!}{\includegraphics{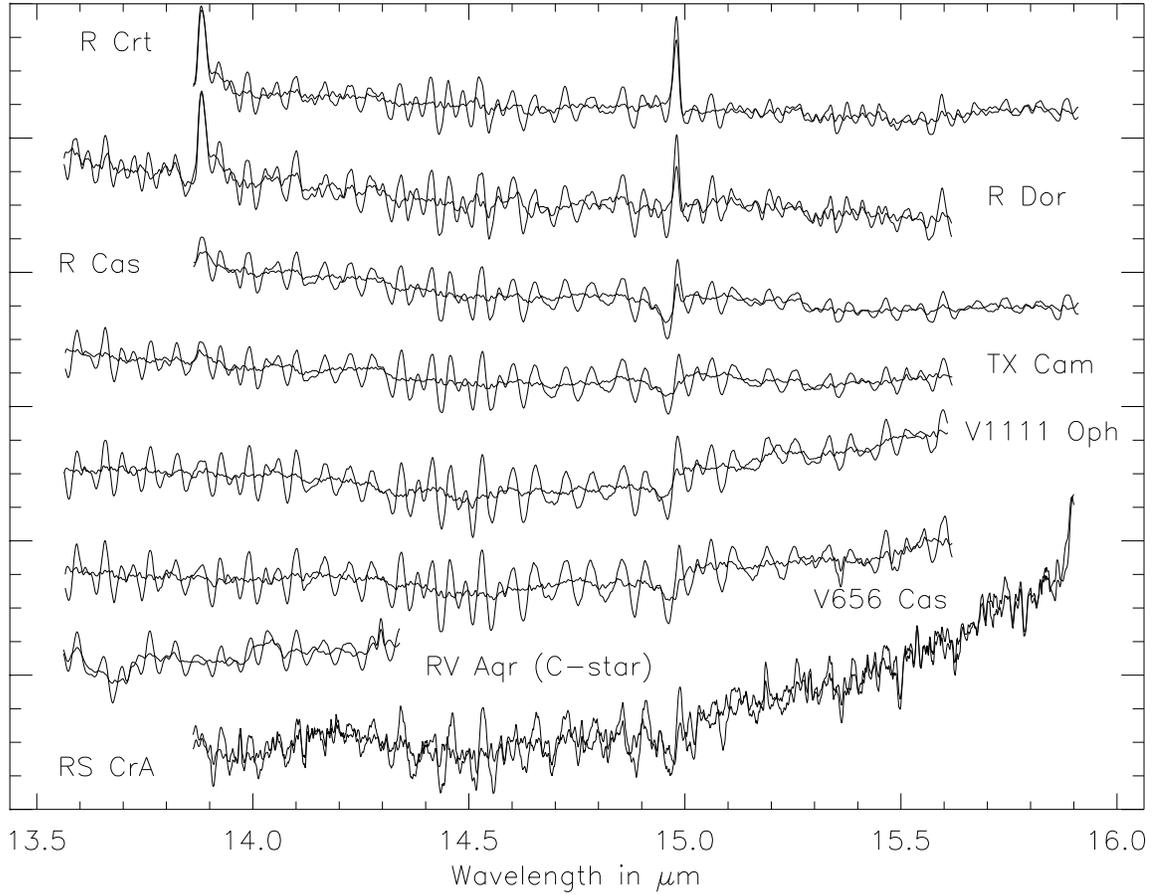}}
  \caption{The seven observed M giants and the carbon star \object{RV Aqr} as observed by ISO. The data are scaled in strength to make a qualitative comparison possible between the stars. Superimposed are the defringed spectra. The observed CO$_2$ Q-bands lie at  $13.9\mbox{$\mu$m}$ and at $15.0\mbox{$\mu$m}$}
  \label{alla}
\end{figure*}

The spectra of the seven M giants and the C star \object{RV Aqr} were observed with the Short-Wavelength Spectrometer (SWS, de Graauw, et al. 1996\nocite{degraauw}) on board ISO \cite{kessler}. The
spectrometer was used in the grating scan mode (SWS06), providing a resolution of ${\rm R\sim 1900}$, corresponding to $\Delta \lambda \sim 8\cdot 10^{-3} \mbox{$\mu$m}$ or  $\Delta v\sim 160\,\mathrm{km/s}$. 
Finally one observation of \object{R Dor} was performed using the high resolution SWS07 mode of 
the Fabry-P\'erot interferometer with $\Delta \lambda \sim 0.5\cdot 
10^{-3}  \mbox{$\mu$m}$. 
The field of view of the SWS06 observation was $14\arcsec$~times~$27\arcsec$. 

\begin{table*}
  \caption{Basic data for M giants with CO$_2$ features, according to Loup et al. (1993)}
  \label{tab1}
  \begin{tabular}{l l l l l l l l | l l} \hline
  \noalign{\smallskip}
  M giant & Type & Class & Period & ${\rm \dot M}$ & v$_{\rm e}$ & $\rm {r_{CO}}$ & Distance & Date of observation & Feature \\
  &  &  & $\mathrm{(days)}$ & $\mathrm {(M_\odot\,yr^{-1}) }$ & $\mathrm {(km\,s^{-1})}$ & $\mathrm {(cm) }$ & $\mathrm {(pc)}$ & \& revolution no.& \\
  \noalign{\smallskip}
  \hline
  \noalign{\smallskip}
  \object{R Crt} & M7III  & SRb  & $160$     & $7.5  \cdot10^{-7}$& $11$ & $ 0.68 \cdot 10^{17}$ & $250$ &  960202 (228) & em \\
  \object{R Dor} & M8IIIe & SRb  & $338$     & $0.7 \cdot 10^{-7}$& $6 $ & $ 0.21 \cdot 10^{17}$ & $40 $ & 970623 (585)  & em \\
& & & & & & & & and 971228 (773)& \\
  \object{R Cas} & M7IIIe & mira & $430$     & $11 \cdot 10^{-7}$ & $12 $& $0.80 \cdot 10^{17}$ & $200$   & 961203 (382) & em (?)\& abs \\
  \object{TX Cam}    & M8.5     & SRb    &  $557$    & $25  \cdot 10^{-7}$& $18$ & $1.2  \cdot 10^{17}$& $330 $ & 970823 (646)& abs \\
  \object{V1111 Oph} & M9       & mira   &         & $60 \cdot 10^{-7}$& $17$ & $1.9 \cdot 10^{17}$ & $570$  & 970424 (525)& abs  \\
  \object{V656 Cas}  & M9       & mira   &         & $69  \cdot 10^{-7}$& $14$ & $2.0 \cdot 10^{17}$ & $950$  &970928 (683)& abs \\
  \object{RS CrA}    &          &        &         & $28 \cdot 10^{-7}$& $21$ & $1.1 \cdot 10^{17}$ & $280$  &  961015 (334)& abs (?) \\
  \noalign{\smallskip}
  \hline
  \end{tabular}
\end{table*}

The data were partly reduced at the ISOSDC (ISO Spectroscopy Data Centre) at  the Max-Planck Institut f\"ur Extraterrestrische Physik in Garching, using the 1997 version of the Interactive Analysis package (IA), and in Uppsala 
using ISAP 1.5 (ISO Spectral Analysis Package). The observed bands are not heavily crowded by cosmic ray hits, but
are unfortunately severely affected by fringes due to regular interference patterns, which depend on the exact pointing of the satellite at the time of observation. Therefore, the fringes can not be reduced away directly. A possible way to artificially eliminate the fringes is, for example, to divide the signal by the simple harmonics that dominate the fringe pattern. This was done by using a defringing algorithm included in the IA package. Such a procedure is, however, somewhat questionable since real features, especially molecular lines occurring at frequencies close to those of the fringe pattern, are also taken away if not special care is taken.  The flux calibration leads to a relative uncertainty of less than $10\%$.

The M giants observed are given in Table \ref{tab1} and the observations of these stars are shown in Fig.~\ref{alla}. The observations are all shown in the same figure and are therefore shifted and scaled arbitrarily in vertical direction to enable a comparison between the stars. Superimposed on the original, fringed data are the artificially defringed data. The carbon star \object{RV Aqr}, supposedly without any CO$_2$ but with the same instrumental fringes, is plotted to give evidence for
the hypothesis that the fringes are instrumental and not specific for M stars. The fringe pattern shows a clear resemblance among the M stars, and is convincingly similar between the M giant spectra and the carbon star spectrum. 
The spectra of \object{R Cas} and \object{R Crt} start from a little less than $13.9\mbox{$\mu$m}$, which means that the $13.9\mbox{$\mu$m}$ band is included but is fairly close to the edge of the detected spectrum. This part of the spectrum is sampled only once, whereas the central regions are covered twelve times. Thus, the edges are more sensitive to cosmic hits and may therefore be noisier.  The spectrum of \object{RS CrA} is of poor quality due to the low flux measured.

\begin{figure}
  \resizebox{\hsize}{!}{\includegraphics{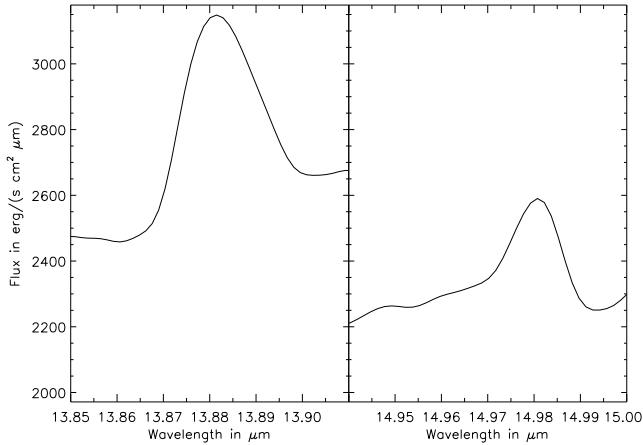}}
  \caption{Details of the defringed spectrum of \object{R Dor}. The CO$_2$ Q-bands at  $13.9\mbox{$\mu$m}$ and at $15.0 \mbox{$\mu$m}$ are asymmetric, the former to the red and the latter to the blue}
  \label{Rdor_intill}
\end{figure}

The CO$_2$ features are found around $13.9$ and $15.0\mbox{$\mu$m}$. 
Visible emission at $13.9 \mbox{$\mu$m}$ is seen in the spectra of \object{R Crt} and \object{R Dor}, and possibly in the spectrum of \object{R Cas}. The feature at $15.0 \mbox{$\mu$m}$ is seen as an emission in the same stars. In \object{R Cas} there is, in addition to the emission feature, an absorption part which is broader. In \object{TX Cam}, \object{V1111 Oph}, and \object{V656 Cas} and possibly \object{RS CrA} the feature is seen mainly as an absorption. In Fig. \ref{Rdor_intill} the $13.9$ and $15.0\mbox{$\mu$m}$ emission bands of \object{R Dor} show asymmetries, the former to the red and the latter to the blue. This is seen in all the stars that show CO$_2$ emission.

\section{The carbon dioxide rotation-vibration bands}

\begin{figure}
  \resizebox{\hsize}{!}{\includegraphics{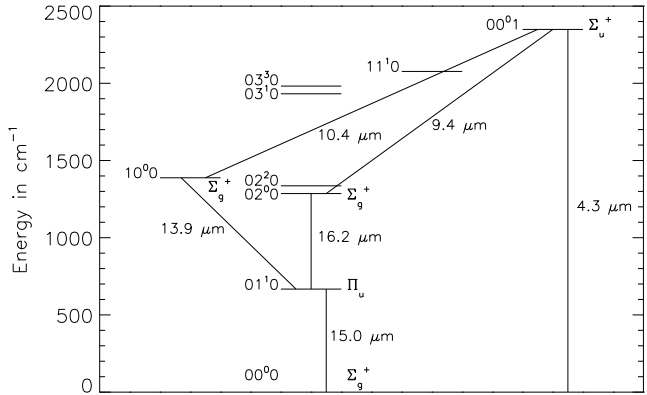}}
  \caption{The vibrational energy-level diagram of some states in CO$_2$. A number of infrared transitions are indicated. The labels identifying the different levels indicate the number of quanta in the three fundamental vibrational modes. The superscript denotes the quantum number associated with vibrational angular momentum, $l$
    }
  \label{levels}
\end{figure}

The $\nu_2$ bending-mode, rotation-vibrational band at 
$667\mbox{cm$^{-1}$}$ or $15.0\mbox{$\mu$m}$, is a $\Sigma_g^{ \,+} - \Pi_u$ transition of the linear and symmetric molecule CO$_2$, see e.g. Herzberg \cite*{herz}. 
The selection rules for this perpendicular band, $\Delta v=\pm 1$ and $\Delta 
J=0, \pm 1$, result in a Q, P, and R-branch.  The Q-branch will appear strongest 
since it will correspond to an overlap of 
$\Delta J=0$ transitions, whereas the P- and R-branches will be smeared out and will hardly be detectable at the resolution used in the observations. If the rotational constant ${\rm B}$, determining the energy
separations of the rotational levels within a vibrational state, would be the same for the upper vibrational state as for the lower one, the $\Delta J=0$ transitions would all have the same frequency. The rotational constant for the upper vibrational level (${\rm B'}$) is in the $15\mbox{$\mu$m}$ band case, however, modestly larger than that of the lower level (${\rm B''}$), leading to a small increase in frequency for high $J$-values. Therefore, the Q-branch is expected to be slightly asymmetric to the blue. For the $13.9$ and $16.2\mbox{$\mu$m}$ bands (cf. Fig.\ref{levels}) ${\rm B' < B''}$ implying a reversed asymmetry. The width of a thermal band reveals the underlying temperature since as the temperature increases, higher rotational levels will be populated, resulting in a broader band. The difference of the rotational constant of the upper and lower levels gives an indication of how sensitive the broadening is to the temperature. 
In Table \ref{tab_einstein} the band oscillator strength, ${\rm f_{band}}$ and the difference between the rotational constants of the upper and the lower level of the transition, ${\rm B'-B''}$, are given for the three bands discussed here.

The $13.9 \mbox{$\mu$m}$ Q-band is a transition from a state where the molecule vibrates in the $\nu_1$-mode and de-excites to the upper level of the $15.0 \mbox{$\mu$m}$ band. The Q-band at $16.2 \mbox{$\mu$m}$ 
is a de-excitation of a molecule vibrating in the second level of the $\nu_2$ mode down to the upper level of the $15.0 \mbox{$\mu$m}$ band. See Fig. \ref{levels}.

The resolution of the measurements ($\Delta \lambda \sim 0.008\mbox{$\mu$m}$) will allow the asymmetry of the observed features (FWHM$\sim 0.015 \mbox{$\mu$m}$) to be measurable, which is also observed. The rotational fine structure, i.e. the individual Q-lines will, however, not be detectable in the SWS06 mode of the spectrometer. In the high resolution mode (SWS07, $\Delta \lambda \sim 0.0005\mbox{$\mu$m}$) the Q-lines will show up as separated symmetric lines, since the typical separation for the lowest Q-lines is approximately $0.001\mbox{$\mu$m}$.

\begin{table}
  \caption{Parameters of the CO$_2$ Q-bands. The band oscillator strengths were derived from Rothman et al. (1992) and the rotational constants are directly given in the same reference}
  \label{tab_einstein}
  \begin{tabular}{l l l } \hline
  \noalign{\smallskip}
  wavelength & ${\rm f_{band}}$ & ${\rm B'- B''/cm^{-1}}$ \\
  \noalign{\smallskip}
  \hline
  \noalign{\smallskip}
  $13.87\mbox{$\mu$m}$ & $ 0.167\cdot 10^{-6} $ & $ -1.06 
\cdot 10^{-3}$\\
  $14.97\mbox{$\mu$m}$ & $ 8.65\cdot 10^{-6}  $ & $ +1.03
\cdot 10^{-3}$ \\
  $16.18\mbox{$\mu$m}$ & $ 0.163\cdot 10^{-6} $ & $ -0.77
\cdot 10^{-3}$ \\
 
  \noalign{\smallskip}
  \hline
  \end{tabular}
\end{table}

\section{Results and analysis}

\nocite{roth92}
Observational evidence points toward a relative simplicity of the kinematics, the density and the
temperature structures of stellar envelopes, with an over-all symmetry prevailing (see e.g. Olofsson, 1996\nocite{ho:rev}), although several AGB stars have been shown to have asymmetric winds (Plez \& Lambert, 1994 \nocite{plez}). Also, note the clear evidence of clumpy structure at some scale in e.g. Olofsson et al. \cite*{ho:98}. 
The chemical structure is certainly quite complex, with a chemistry far out of equilibrium.
A standard circumstellar envelope, with a non-rotating, smooth, and spherically expanding wind with a constant velocity is nevertheless a reasonable starting point for our analysis.

\subsection{Are the bands optically thin?}

In the following we will discuss the optical thickness of the 
environment from which the bands are emitted.  The order-of-magnitude estimate made here will lead to an upper limit partly because the wind velocity will eventually shift the molecules out of the thermal broadening of the Q-branch lines due to Doppler shifts and thereby lower the estimated optical depth.
For an estimate of this upper limit of the optical depth, $\tau$,  of the bands, we assume a spherically symmetric wind resulting from a constant rate of mass-loss.  Within the region where CO$_2$ is assumed to exist, 
its abundance relative to hydrogen, $\epsilon_{_{\mathrm{CO_2}}}$, is set constant.
The optical depth is found by integrating the cross section per 
particle ($\sigma_{\lambda}$, [cm$^2$]) for the absorption times the number density of 
the absorber ($N$, [cm$^{-3}$]) along the line of sight with an impact 
parameter $p$.
For atomic or molecular scattering $\sigma_{\lambda} = \pi \mathrm {e^2} \lambda^2 \mathrm{f_{band}/ (m_e c^2 }\delta \lambda) $, with f$_{\rm{band}}$ being the band oscillator strength. 
From the continuity equation and the assumption of a stationary flow the number density of CO$_2$ in all energy states 
is found.
Thus, in cgs units, 
\begin {equation}
\tau (p) = 21\cdot 10^9 
\epsilon_{_{\mathrm{CO_2}}} \frac{\dot M \lambda^2 \mathrm{ f_{band}}}{v_\mathrm{e} \delta \lambda} \frac{2}{p} 
\arctan ( \frac{ \sqrt{R_{\mathrm{o}}^2-p^2} }{p} ),
\end{equation}
 where $R_\mathrm{o}$ is the outer boundary 
of the emitting sphere and $\mathrm{f_{15 \mu 
m}=9\cdot10^{-6}}$.
The largest optical depth will be found in the $15.0\mbox{$\mu$m}$ band since the 
lower lying energy state of this transition is the vibrational ground state, 
where all but a few molecules will most probably be found. 
An upper limit of the optical depth will therefore be estimated for an impact 
parameter $p=1\,\mathrm{R_*}$ and assuming that all CO$_2$ is in the ground 
state.
An upper limit of the fraction of CO$_2$ ($\epsilon_{_{\mathrm{CO_2}}}$) is found by adopting a reasonable number of the CO$_2$/CO ratio and 
assuming that almost all carbon is locked in CO. This is not a bad assumption in an 
oxygen-rich environment since the CO molecule has a large binding energy 
and is the most stable molecule in the circumstellar environment. 
The CO$_2$/CO ratio is 
assumed be a few times $10^{-4}$, in accordance with the calculated abundances in the 
circumstellar chemical models of Willacy \& Millar \cite*{willacy}. The width 
of the band ($\delta \lambda$) is approximately $0.02\mbox{$\mu$m}$.

The results for the stars considered are tabulated in Table \ref{opt_djup}, which gives the  optical depths for a standard wind extending out to around 1000 stellar radii. Obviously, all bands will be optically thin in this case. A wind extending out only to 3 stellar radii, would give around 20\% lower optical depths under the same assumptions. If we consider a warm and quasi-static, molecular-forming region with small velocities, such as the one suggested by Tsuji et al. \cite*{tsuji}, the densities close to the star are enhanced,
which will lead to higher optical depths.  A density enhancement close to the star by one or two orders of magnitude could make the $15 \mbox{$\mu$m}$ band optically thick. An increase of the density could be achieved if extra ejected matter eventually falls back on the star without raising the net matter transport outwards. Turbulent velocities will broaden the lines and the quasi-static layer  will not shift the molecules out of the lines. 
Thus, there is a possibility that the absorption bands observed are formed in such high density regions.
In view of their lower oscillator strengths and their excitation, 
the other bands are obviously optically thin also in this picture.

\begin{table}
  \caption{Maximum optical depths,${\rm \tau_{max}}$, for the $15.0 \mbox{$\mu$m}$ band for a spherical and stationary wind}
  \label{opt_djup}
  \begin{tabular}{l l} \hline
  \noalign{\smallskip}
  M giant & maximum optical depth, ${\rm \tau_{max}}$  \\
  \noalign{\smallskip}
  \hline
  \noalign{\smallskip}
  \object{R Crt} & $0.07$\\
  \object{R Dor} & $0.01$\\
  \object{R Cas} & $0.09$\\
  \object{TX Cam}    & $0.13$\\
  \object{V1111 Oph} & $0.33$\\
  \object{V656 Cas}  & $0.05$\\
  \object{RS CrA}  & $0.13$\\
  \noalign{\smallskip}
  \hline
  \end{tabular}
\end{table}

\subsection{Are the bands formed in LTE? \label{LTE}}

It is widely believed that most AGB stars pulsate with shocks travelling through their 
circumstellar regions. These dynamic regions are often assumed to show departures from 
LTE conditions. We will show that it may be questionable to assume that the bands are in LTE.

The highest probability for LTE to prevail, in an expanding wind model, is close to the star since the density 
decreases, at stationary outflow, with the square of the distance from the centre. From the equation of 
continuity for a spherically symmetric wind we get the number density of 
hydrogen molecules as a function of the radial distance from the star r, as 
$N_{_{\mathrm{H_2}}} = \dot M/(4\pi v_\mathrm{e} r^2  \mathrm{m_{_{H_2}}}).$ For \object{R Crt} with 
$\dot M=8\cdot 10^{-7}\, \mathrm{M_{\sun}\,yr^{-1}} $ and $v_\mathrm{e}=11\, \mathrm{km\,s^{-1}}$ and for \object{R Dor}  
with $\dot M=7\cdot 10^{-8}\,\mathrm{ M_{\sun}\,yr^{-1}}$ and $v_\mathrm{e}=6\, \mathrm{km\,s^{-1}}$ \cite{loup}, we get for the number density of H$_2$ for this stationary model, adopting $\mathrm {R_*=3\cdot 10^{13}\mbox{cm}}$:  
$1\cdot 10^{9}\cdot (R_*/r)^2\mbox{cm$^{-3}$}$ and $0.2\cdot 10^{9}\cdot (R_*/r)^2\mbox{cm$^{-3}$}$, 
respectively. 

Observe, however, that the atmospheres of these stars extend far out due to pulsations and shocks, which will invalidate the assumption that the wind is in a steady state at small radii. 
Thus, from dynamic models of variable stars by, e.g., Bessel et al. \cite*{bessel} and H\"ofner \& Dorfi \cite*{dorfi}, it is indeed seen that the densities at a few stellar radii from the star can be, in general, an order of magnitude higher than the ones given by the equation of continuity. Note also that observations indicate that the winds may be inhomogeneous and clumpy (e.g. Olofsson et al., 1998\nocite{ho:98}), which could increase the densities in the emission regions. Also the quasi-static layer discussed above will have higher densities, which could be enough for collisions to be dominant a few stellar radii out.  However, from the simple considerations above we will see that we may nevertheless conclude that the assumption of statistical equilibrium may be needed instead of that of LTE, even a few stellar radii away from the star.
This must be considered when discussing the kinetic gas temperature and the site of formation of CO$_2$.

The  statistical equilibrium equations, including radiative and collisional processes for a set of transitions, give the populations of the levels. These will also be given by the Boltzmann distribution if the density is sufficiently high in order for collisions to be dominant over radiation processes. The spectral lines are then formed in LTE and the gas is 
thermalised at the kinetic gas temperature.  If the collisions are not dominant, it is not clear whether LTE is valid.

An order-of-magnitude estimate of the probability for collisional excitation is given by $\mathrm{C_{lu}}\sim N_{_{\mathrm{H_2}}}\sigma\,v $.
The cross section, $\sigma$, can be approximated with the geometrical area of a molecule, 
which is about $\sigma_{\mathrm{geom}}=10^{-16}\mbox{cm$^{2}$}$.
The velocities of the 
molecules for the temperatures considered are of the order of 
$v\sim 1\mbox{km\,s$^{-1}$}$.
For the Q-band at $15.0\mbox{$\mu$m}$, with an 
Einstein coefficient for spontaneous decay of  $\mathrm{A_{ul}}\sim 1\mbox{s$^{-1}$}$, the probability of radiative excitation  ($\mathrm{ B_{lu}}\cdot J_{\nu}$, $J_{\nu}$ being the mean intensity of the star) is $0.3\mbox{s$^{-1}$ }$ for a typical M giant and extrapolating back to $\mathrm{1\,R_*}$. In order for the collisional processes to dominate
we thus need $N_{_{\mathrm{H_2}}}\gg 3\cdot 10^{10}\mbox{cm$^{-3}$}$.
Therefore, it seems that the densities in the circumstellar environments, 
are not high enough for the collisions to be completely dominant. Radiative processes may therefore be more important than collisional ones, which 
may lead to severe departures from LTE. 

However, among the rotational energy levels in the vibrational 
ground state the levels are thermally populated. This conclusion can be drawn 
since the Einstein coefficient for spontaneous decay between the rotational 
states are several orders of magnitude smaller than the coefficients for vibrational decays. 
The relative population of the levels within the vibrational ground state are therefore given by the  Boltzmann distribution and the local gas temperature. The population of the rotational levels of higher lying vibrational states are affected by radiative transitions between the different vibrational states.

In a situation dominated by radiative processes, the strongest transitions involved in a case of statistical  equilibrium are shown in Fig. \ref{levels}.  These levels and the transitions between them were included in a Monte Carlo simulation of a radiative situation to discern to what extent the transitions will eventually spread out the populations of the rotational levels within higher lying vibrational states. In this situation the rotational levels of the ground state  are given by the Boltzmann distribution and the probabilities of the transitions, i.e. the oscillator strengths, of an individual, rotation-vibration transition within a band are given by

\begin{equation}
\label{f_line}
\mathrm{f^{abs}_{line}}=\frac{ \mathcal{S}_{J{''}J'}}{2J{''}+1} \mathrm{f_{band}} \left( \frac{\lambda_{\mathrm{centre}}}{\lambda_{\mathrm{line}}} \right) \frac{2-\delta_{0,\Lambda{''}}}{2-\delta_{0,\Lambda{'}+\Lambda{''}}},
\end{equation}

where $\Lambda$ denotes the projection of the electronic orbital angular momentum along the axis of the linear molecule and $\mathcal{S}_{J{''}J'}$ are the H\"onl-London factors which can be found in, e.g. Bernath \cite*{bernath}. These can be calculated for all the transitions considered.

The result of this investigation shows that the spreading out of the level populations is on a few percent level and hardly noticeable. This means that also in an environment dominated by radiative processes the relative rotational populations within the higher lying vibrational levels will map the Boltzmann distribution of the ground vibrational level. This leads to the conclusion that the widths of all vibration bands will reflect the underlying gas temperature. However, in general, the relative strengths or amplitudes of the bands will {\it not} be given by the thermal distribution for the vibrational levels, if the radiation processes dominate. The $13.9$ and $16.1\mbox{$\mu$m}$ bands originate from upper states of about the same energy and they are affected similarly by radiation excitation, which may, in any case, lead to similar band strengths.

\subsection{The temperature of the bands}

\begin{figure}
  \resizebox{\hsize}{!}{\includegraphics{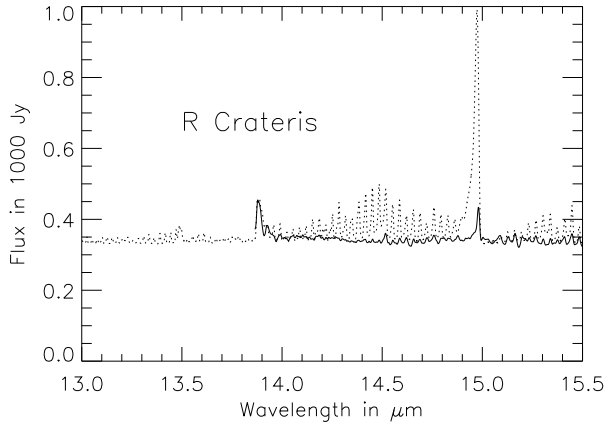}}
  \caption{The full line shows the two observed Q-bands of \object{R Crateris} and the dashed line represents a thermal 
model of $650\mbox{K}$ scaled to the $13.9\mbox{$\mu$m}$ transition}
  \label{fig1}
\end{figure}

\begin{figure}
  \resizebox{\hsize}{!}{\includegraphics{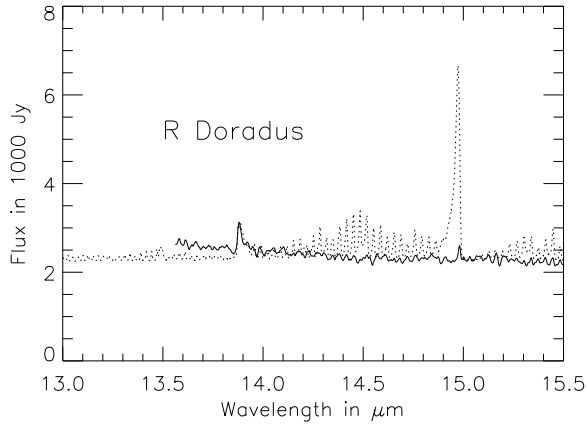}}
  \caption{As in Fig.\,\ref{fig1} but for \object{R Doradus}. The forests at $14.5$ and $15.5\mbox{$\mu$m}$ are the R and P branches of the 
$15.0\mbox{$\mu$m}$ transition in the model of $650\mbox{K}$. These lines are not resolved in the resolution of the observation}
  \label{fig2}
\end{figure}

In this section we account for the excitation temperatures of the different bands for the  stars showing emission. This is done by modelling the band widths of a thermally populated CO$_2$ molecule and fitting its spectrum to the observed bands. The HITRAN molecular database \cite{HITRAN} and the high temperature analogue HITEMP database (Rothman et al., in prep.) are used to retrieve relevant spectral line data such as line positions, level populations and transition probabilities.

The widths of the bands depend on the population of the rotational levels within the upper vibrational state and thereby the excitation temperature. In a thermal model these are populated according to the Boltzmann distribution.

The observations of the $15.0$ and the $13.9\mbox{$\mu$m}$ Q-bands of carbon dioxide in \object{R Dor} and \object{R Crt} are shown in Figs.~\ref{fig1} and \ref{fig2}. These spectra show the residual emission after the fringes, due to the spectrometer, have been removed.
Also plotted in the figures are the bands of a thermal model for $650\mbox{K}$ smoothed to the resolution of the observations. 
The emission predicted by the model is plotted with the observed continuum as a base line and is  
scaled  to fit the $13.9\mbox{$\mu$m}$ band. 

For \object{R Dor} the excitation temperature as estimated from the $13.9\mbox{$\mu$m}$ band width is  $T_{\mathrm{ex}}=(650\pm 50)\mbox{K}$, while the $15.0\mbox{$\mu$m}$ band is fitted best with $T_{\mathrm{ex}}=(150\pm 50)\mbox{K}$. The band ratio between the $13.9\mbox{$\mu$m}$ and the $15.0\mbox{$\mu$m}$ bands is  $1.7$.

\begin{figure}
  \resizebox{\hsize}{!}{\includegraphics{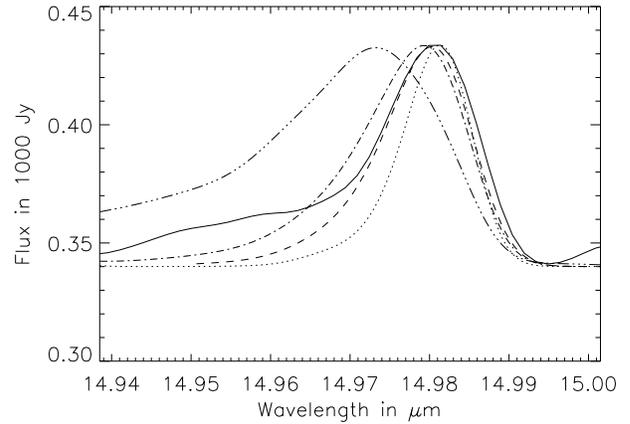}}
  \caption{The observed $14.98\mbox{$\mu$m}$-band for \object{R Crt} is plotted in full line. Arbitrarily scaled thermal models for temperatures of $125\mbox{K}$, $160\mbox{K}$, $250\mbox{K}$, and $650\mbox{K}$ are also plotted in the figure. The maximum of the model spectra is shifted towards the blue with increasing temperature. The relative errors in the observations are on the order of a few percent}
  \label{160K}
\end{figure}

Also for \object{R Crt} the $13.9\mbox{$\mu$m}$ band suggests $T_{\mathrm{ex}}=(650\pm 50)\mbox{K}$, whereas the $15.0\mbox{$\mu$m}$ band is better fitted with a thermal model with a temperature of $T_{\mathrm{ex}}=(160\pm 50)\mbox{K}$, cf. Fig. \ref{160K}. The observed band ratio is for this star $1.0$.

\begin{figure}
  \resizebox{\hsize}{!}{\includegraphics{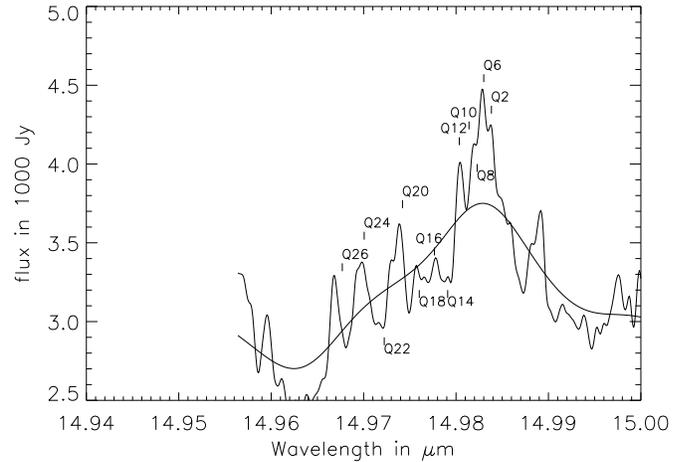}}
  \caption{The high-resolution Fabry-P\'erot observations of \object{R Dor}. These observations are very difficult to reduce due to 
cosmic ray hits, and spurious features are a problem. Also plotted in the figure is the spectrum smoothed to the resolution corresponding to the other observations}
  \label{FP}
\end{figure}

It is also clear from Figs. \ref{fig1} and \ref{fig2} that the band intensity
ratios can not be explained by a simple, optically thin, one-component LTE-model.  Our observations of the $15.0\mbox{$\mu$m}$ band of \object{R Dor} in the high resolution mode also show the 
individual Q-lines. These observations are unfortunately very difficult to reduce due to 
cosmic hits. However, from the observations, it can be concluded that the intensity
distribution of the different rotational lines within this vibrational band suggest a temperature several hundred degrees below $650\mbox{K}$, cf. Fig. \ref{FP}. At $650\mbox{K}$ the maximum line strength lies around the transition Q24, and at $150\mbox{K}$ at Q10.
Also, the Fabry-P\'erot observations smoothed to a resolution corresponding to the one used in the other spectra, can not be fitted with a $650\mbox{K}$ model, cf. Figs. \ref{160K} and \ref{FP}.

\subsection{The number of emitting molecules \label{sect}}

\begin{table}
  \caption{Number of excited CO$_2$ molecules required for the measured flux and the total number of molecules for two different shell sizes. See text}
  \label{CSE}
  \begin{tabular}{l l l l l } \hline
  \noalign{\smallskip}
  M giant & ${\rm Flux/ }$ & $\rm{N_{CO_2}^*/}$ & $\rm{N_{CO_2}^{tot}/}$  & $\rm{N_{CO_2}^{tot}/}$  \\
   & $10^{-11}$ &   $\rm{10^{44}}$ &  $\rm{10^{44}}$ & $\rm{10^{44}}$  \\
   & ${\rm erg\,s^{-1}\,cm^{-2}}$ &  & $\rm{(3\,  R_*)}$  & $\rm{(large\, \, shell)}$  \\
  \noalign{\smallskip}
  \hline
  \noalign{\smallskip}
  \object{R Crt} & & & & \\
  $13.9\mbox{$\mu$m}$  & $\rm{3.9\pm 0.4}$  & $\rm{ 350\pm 40 }$  & $\rm{ 0.6}$ &  \\
  $15.0\mbox{$\mu$m}$  & $\rm{2.9\pm 0.3}$  & $\rm{ 12\pm  1 }$  & $\rm{ 0.6}$ & $\rm{ 300}$ \\
  \object{R Dor} & & & & \\
  $13.9\mbox{$\mu$m}$  & $\rm{21\pm 2}$     & $\rm{ 48\pm 5 }$  & $\rm{ 0.09}$ &  \\
  $15.0\mbox{$\mu$m}$  & $\rm{9.5\pm 1}$    & $\rm{ 0.97\pm 0.1 }$  & $\rm{ 0.09}$ & $\rm{ 50}$ \\
  \object{R Cas} & & & & \\
  $15.0\mbox{$\mu$m}$  & $\rm{2.3\pm 0.3}$    & $\mathrm {5.9\pm 0.6}$ &  & $\rm{ 400}$ \\
  \noalign{\smallskip}
  \hline
  \end{tabular}
\end{table}

The number of molecules needed in a certain vibrational state in order for an emitting optically thin envelope to 
radiate a given measured flux, $F_{\mathrm{m}}$, can be calculated. With the Einstein coefficient for spontaneous decay, ${\rm A_{ul}}$, the total number of molecules in the upper level, $N_{\mathrm{tot, u}}$, is given by 
\begin{equation}
\label{n_tot}
N_{\mathrm{tot, u}}=\frac{4\pi \cdot \mathrm{distance}^2 \cdot F_{m}}{\mathrm{h}\nu \mathrm{A_{ul}}}.
\end{equation}

This number should be compared to the total number of available CO$_2$ molecules  in all vibrational states, which 
can be estimated 
from
\begin{equation}
N(\mathrm{CO_2})=\frac{\dot M}{v_\mathrm{e}\, m_{_{\mathrm{H_2}}}}(R_{\mathrm{o}}-R_{\mathrm{i}})\,\epsilon_{_{\mathrm{CO}}}\,\mathrm{\frac{[CO_2]}{[CO]}},
\end{equation} 
where $\mathrm {\epsilon_{_{CO}}}$ is the abundance of CO relative to hydrogen, $\dot M $ the mass-loss rate, $v_{\mathrm{e}}$ the terminal wind velocity and 
$R_{\mathrm{o}}$ and $R_{\mathrm{i}}$
are the outer and inner radii of the emitting shell, respectively.
As before, we adopt a number ratio of CO$_2$ 
molecules to CO molecules of a few $10^{-4}$.

Using these expressions we can compare the number of excited CO$_2$ molecules needed for the measured flux to the total amount of CO$_2$. The measured fluxes in the $13.9$ and $15.0\mbox{$\mu$m}$ bands are presented in Table \ref{CSE}. The table also gives the number of excited molecules required for the measured flux as well as the total number of molecules in an emitting shell, 3 stellar radii wide, and a large emitting envelope. The extension of the CO$_2$ shells for \object{R Dor}, \object{R Crt} and \object{R Cas} used for the results in the last column in Table \ref{CSE}, are found by scaling the results for \object{TX Cam} from the models of Willacy\,\&\,Millar.
This column therefore gives the number of carbon dioxide molecules in a large emitting shell extending from 50 to 1000 R$_*$ for \object{R Crt} and \object{R Cas} and  from 20 to 300 R$_*$ for \object{R Dor}.

\section{Discussion}

The excitation temperatures derived from the different bands are not in agreement with each other. We therefore suggest a two component model which could explain the spectra.

In the first component, corresponding to the region suggested by Justtanont et al. \cite*{jus}, the CO$_2$ exists in a dense quasi-static layer close to the star, the layer extending out to only a few stellar radii, resembling the warm, molecular-forming region suggested by Tsuji et al. \cite*{tsuji}. CO$_2$ is supposed to have a temperature of the order of  $650\mbox{K}$ and high densities.
This would give rise to the $13.9\mbox{$\mu$m}$ band and the absorption at $15\mbox{$\mu$m}$, a band which is now optically thick and which will appear differently depending on the physical conditions. As seen in Table \ref{CSE}, the number of molecules needed for the observed emission is too high compared to the number calculated from a  spherically symmetric wind at a constant mass-loss rate. It is impossible to produce these fluxes from a small region close to the star extending out to, e.g., 3 stellar radii with the assumptions above. The number of excited molecules needed exceeds the expected number for a thermally populated level by several orders of magnitude. 

More reasonable numbers will result from a high density layer with a density of more than $N_{\mathrm{H_{_2}}}=10^{10}\mbox{cm$^{-3}$}$, only  extending out to a few stellar radii. A density enhancement of several orders-of-magnitude above the classical wind density, is required. In this case, the $15\mbox{$\mu$m}$ band will be formed in conditions in which collision are close to being dominant over radiative processes, cf. Sect. \ref{LTE}.

The second component, which has been argued for  by Ryde et al. \cite*{ryde97}, would be a region extending far out from the star. The colder $15\mbox{$\mu$m}$ emission is then supposed to originate from this tenuous region. The corresponding cold $13.9\mbox{$\mu$m}$ emission would  be hardly detectable in the $650\mbox{K}$ emission.  
Thus, as is seen in Table \ref{CSE}, assuming a much larger shell of CO$_2$, extending out to several hundred stellar radii (finally limited by the photo dissociation by the interstellar UV field) the required degree of excitation in the upper level of the 
$15.0\mbox{$\mu$m}$ band  would be some percent for all our stars, which is not unreasonable. The number of stellar IR photons per square centimetre and second is typically two orders of magnitude greater than the number of photons in the measured lines, which is reassuring.

The temperature and number of molecules required from the observed $15\mbox{$\mu$m}$ emission features  are thus consistent with the location of the carbon dioxide  in a circumstellar shell, extending typically out to $10^3$ stellar radii. This is also consistent with the chemical models of Willacy \& Millar (1997) for oxygen-rich circumstellar envelopes, in which CO$_2$ is formed in a region 
from around 100 to 1000 stellar radii. Abundances of other molecules in their models show good agreement with observations, which is reassuring. In these models carbon is provided by CO and CH$_4$, and most carbon-bearing molecules are due to the breakdown of CH$_4$. An important reaction for the present discussion is the one where some CO is destroyed at small radii by the reaction CO+OH$\rightarrow$CO$_2$+H. Further out, all molecules will eventually be destroyed by photo-dialysis by the interstellar ultraviolet field.

\object{R Cas} is a special case (Fig. \ref{alla}). Here, both absorption and emission are clearly seen. With the emission from the cold, second component filling in the absorption  at $15\mbox{$\mu$m}$ in the red wing, the width of the absorption is consistent with $650\mbox{K}$. The residual emission in the $15\mbox{$\mu$m}$ band, after the absorption has been subtracted, is $(2.3\pm 0.3)\cdot 10^{-11}\mbox{erg\,s$^{-1}$\,cm$^{-2}$}$, which requires $(5.9\pm 0.6)\cdot 10^{44}$  molecules. This is a reasonable number of excited molecules compared to the total number expected in a large shell, cf. Table \ref{CSE}.

As seen from Table \ref{tab1}, where the parameters of the stars discussed are given, the only parameter that is clearly different between the M giants with and without CO$_2$ emission is the mass-loss rate. In the latter stars the high mass-loss rate seems to prevent the formation of the CO$_2$ emission. This finding was also hinted at by Justtanont et al. \cite*{jus}. They found that the equivalent widths of the CO$_2$ feature are well and positively correlated with the $13\mbox{$\mu$m}$ dust feature. They indicate that the higher the mass-loss rate the smaller is the $13\mbox{$\mu$m}$ feature and thereby the less pronounced are CO$_2$ features. This is consistent with our findings. Whether the correlation between mass-loss rate and the CO$_2$ features has a simple physical origin is still uncertain. It seems, however, tempting to suggest that the stars with high mass-loss rates more easily accelerate winds from the stellar surface to velocities above the velocity of escape while stars with lower mass-loss rates produce less acceleration. For these stars much of the material levitated above the photosphere is stopped by gravity or shocks close to the star where it forms a relatively dense warm layer, where $\mathrm{CO_2}$  and dust may form and radiate.

The relative strengths of the $15.0\mbox{$\mu$m}$ band emission and the $13.9\mbox{$\mu$m}$ band emission are different in \object{R Crt} and \object{R Dor}. This difference in relative strengths is also seen in the six stars of Justtanont el al. \cite*{jus}.
This would be a natural consequence of the two component model. In an optically thin, one component model all molecules de-exciting from the $10^00$ and $02^00$ energy states will finally have to de-excite via the $15.0 \mbox{$\mu$m}$ transition.   This does not seem to occur according to the observations. A two component model with one of the bands being optically thick may solve this. It could also be explained 
if 
the resonance band is on the verge of becoming optically thick, thereby being weakened by self absorption. But in this case one would sooner expect a broader band which is not seen in the observations. 
Our high 
resolution observations also do not verify this.

\section{Conclusions}
The temperatures deduced appear not to be consistent with a model of one single temperature component. Therefore, we have suggested that the bands originate from two different components: One would be a warm and high density layer close to the star, possibly making the $15\mbox{$\mu$m}$ band optically thick, another would consist of a large, cold and optically thin CO$_2$ region extending out in the wind to hundreds of stellar 
radii. These two components could explain the differences in temperatures and in the number of molecules found from the observations. 

The circumstellar CO$_2$ emission bands observed in  
M giants are probably formed out of LTE and caution 
should be exercised when discussing the gas temperature and the site of 
formation of CO$_2$  from observations of the width of an unresolved Q-band.  However, in analysing our seven ISO spectra of M giants, we find that the gas temperatures may be estimated from the widths of the bands even if 
the bands are not in vibrational LTE. We note that Onaka et al. \cite*{onaka} show that all bands correlate with variability phase. This could be explained by differences in the radiation field if the 
population is determined by radiative processes.

Since the CO$_2$ molecules may be dominated by 
radiative pumping by stellar flux via one or some of the many vibrational 
levels, a detailed treatment 
of CO$_2$ in the radiation field and wind of the star seems important for modelling 
the emission and verifying the hypothesis of its origin.

\begin{acknowledgements}

We should like to thank H.\,Olofsson, \& M.\,Asplund for valuable comments and discussions,  D.\,Lutz, D.\,Kunze, \& E.\,Sturm for enlightening discussions, and the Swedish Nat$\rm{^{al}}$
Space Bd. and the Royal Swedish Acad. of Sciences for grants.
Support from the ISO Spectrometer Data Centre at MPE Garching, 
funded by DARA under grant 50 QI 9402 3 is acknowledged.
\end{acknowledgements}

\bibliographystyle{astron}
\bibliography{mnemonic,Ref3}
\end{document}